\documentclass[12pt]{article}
\usepackage{epsfig}
\usepackage{graphicx}
\newcommand{\be}{\begin{equation}}
\newcommand{\ee}{\end{equation}}
\newcommand{\bea}{\begin{eqnarray}}
\newcommand{\eea}{\end{eqnarray}}
\newcommand{\norsl}{\normalsize\sl}
\newcommand{\norsc}{\normalsize\sc}

\newcommand{\nn}{\nonumber}

\def \ksl {k \kern-.45em{/}}
\def \lsl {l \kern-.45em{/}}
\def \ppsl {p \kern-.45em{/}}
\def \qsl {q \kern-.45em{/}}

\begin{document}

\begin{titlepage}

\title{ The  polarized  photon 
structure function $g_1^\gamma(x,Q^2)$ in massive parton model in NLO}
\author{
\norsc Norihisa Watanabe$^{a)}$\thanks{e-mail address: watanabe-norihisa-vz@ynu.ac.jp}, Yuichiro Kiyo$^{b)}$\thanks{e-mail address: ykiyo@tuhep.phys.tohoku.ac.jp}, Ken SASAKI$^{a)}$\thanks{e-mail address: sasaki@ynu.ac.jp}~  \\
\norsl a) Dept. of Physics,  Faculty of Engineering, Yokohama National
University \\
\norsl  Yokohama 240-8501, JAPAN \\
\norsl b)~ Dept. of Physics,  Faculty of Science,  Tohoku University \\
\norsl     Sendai 980-0845, JAPAN \\
}

\date{}
\vspace{2cm}
\maketitle

\vspace{2cm}

\begin{abstract}
{\normalsize
We investigate the one-gluon-exchange  ($\alpha \alpha_s$) corrections to the  polarized  real photon 
structure function $g_1^\gamma(x,Q^2)$ in the massive parton model. 
 We employ a technique based on the Cutkosky rules  and the reduction of 
Feynman integrals to master integrals. The NLO contribution is noticeable at large $x$ and does not vanish at the threshold of the massive quark pair production due to the Coulomb singularity. It is found that the first moment sum rule of $g_1^\gamma$ is satisfied up to the NLO. 
}
\end{abstract}

\begin{picture}(5,2)(-290,-550)
\put(2.3,-95){YNU-HEPTh-11-101}
\end{picture}

\thispagestyle{empty}
\end{titlepage}
\setcounter{page}{1}
\baselineskip 18pt

The experiments at the Large Hadron Collider (LHC) have started and it is much  
anticipated that signals for the Higgs boson and also for the new physics beyond the Standard Model (SM) will be discovered~\cite{LHC}. 
Once these signals are observed, more precise measurements will need to be performed at the future $e^+e^-$  collider, so-called the International Linear Collider (ILC)~\cite{ILC}. In such cases,  a detailed knowledge of the SM at high energies, especially based on QCD, is still important.  

It is well known that, in high energy $e^+e^-$ collision experiments, the cross section of the two-photon processes $e^+e^-\rightarrow e^+e^- + {\rm hadrons}$~ dominates over other processes such as the annihilation process $e^+e^-\rightarrow \gamma^* \rightarrow {\rm hadrons}$. The two-photon processes  at high energies provide a good testing ground for studying the predictions of QCD. 
In particular, the two-photon
process in which one of the virtual photon is very far off shell (large $Q^2\equiv -q^2$), while the other 
is close to the mass shell (small $P^2\equiv -p^2$),  can be viewed as a deep-inelastic electron-photon
scattering where the target is a photon rather than a nucleon.
In this deep-inelastic scattering off a photon target, we can study the photon structure 
functions, which are the analogs of the nucleon structure functions. When  polarized  beams are used in $e^+e^-$ collision experiments, we can 
get  information on  the spin structure of the photon. 

For a real photon ($P^2=0$) target, there exists only one spin-dependent structure function $g_1^\gamma (x,Q^2)$, where $x=Q^2/(2p\cdot q)$.  The photon structure functions are 
 defined in the lowest order of the QED coupling constant 
$\alpha=e^2/4\pi$ and they are of order $\alpha$.  The QCD analysis of $g_1^\gamma$ was performed in the leading order (LO) (the order $\alpha$) \cite{KS}, and in the next-to-leading order (NLO) 
(the order $\alpha\alpha_s$) \cite{SV}, where $\alpha_s=g^2/4\pi$  is the QCD coupling constant. 
In these analyses all the active quarks are treated as massless. 
At high energies the heavy charm and bottom  quarks  also contribute to the photon structure functions. The NLO QCD corrections due to heavy quarks have been 
calculated for the unpolarized photon structure functions  $F_2^\gamma(x,Q^2)$ and 
$F_L^\gamma(x,Q^2)$~\cite{SmithvanNeerven}. 
The heavy quark mass effects on $g_1^\gamma$ were analysed at NLO in QCD in Ref.\cite{GRS} 
by using the LO result of the massive parton model (PM). But the complete heavy quark mass 
effects have not yet been computed for $g_1^\gamma$ at NLO.

In this paper we  investigate the real photon structure function $g_1^\gamma$ in the massive PM at NLO in QCD. In order to compute $g_1^\gamma$ at  NLO,
we employ a technique based on the Cutkosky rules \cite{Cutkosky} and the reduction of 
Feynman integrals to master integrals.
The master integrals which appear in this analysis  
also show up in computing other photon structure functions such as $F_2^\gamma(x,Q^2)$ and 
$F_L^\gamma(x,Q^2)$ at NLO. We express the phase space integrals of these master integrals 
in analytical form as much as possible so that they may serve as useful tools for the analyses
of the future ILC physics.

The polarized real photon structure function $g_1^\gamma$ satisfies a remarkable 
sum rule \cite{ET,BASS,NSV,FS,BBS}
\be
\int_0^1g_1^\gamma(x,Q^2)dx=0~.\label{SumRuleg1gammaReal}
\ee
In particular, applying the
Drell-Hearn-Gerasimov sum rule~\cite{DHG} to the case of a  virtual photon target 
and using the fact that the photon has zero anomalous magnetic moment,
the authors of Ref.~\cite{BBS} 
argue that the sum rule (\ref{SumRuleg1gammaReal}) holds to all orders  
in perturbation theory in both QED and QCD. 
We examine whether the NLO result of $g_1^\gamma$ in the massive PM satisfies this sum rule. 
We find numerically that the sum rule (\ref{SumRuleg1gammaReal}) is indeed satisfied at this order. 
But we point out that the sum rule may not be well-defined when $g_1^\gamma$ is analysed 
to higher orders in perturbation theory, since the calculated result may diverge at 
the threshold of the massive quark pair production due to the Coulomb singularity.

\begin{figure}
\begin{center}
\includegraphics[scale=0.7]{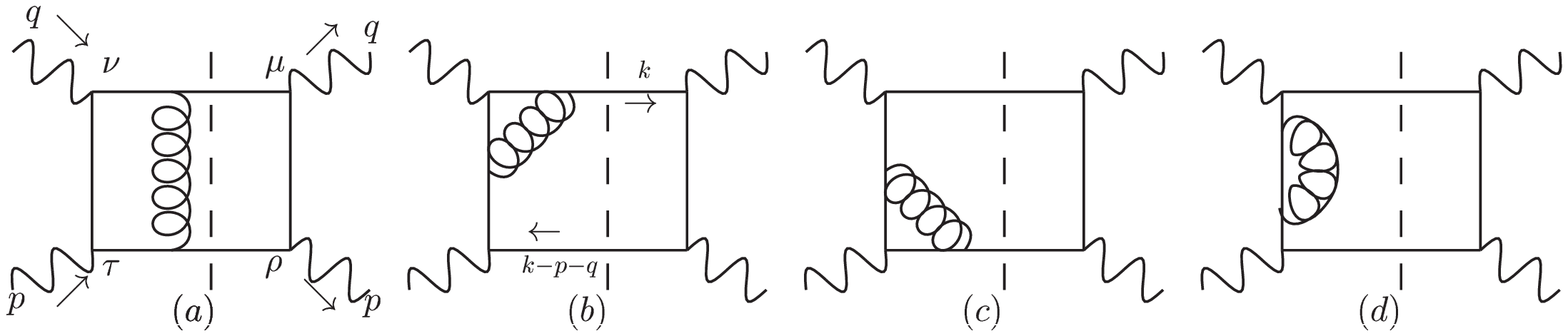}
\caption{Two-loop diagrams with virtual corrections. Graphs with virtual corrections to the right of the cut lines and graphs with $(q, \mu)$ and $(p,\rho)$ interchanged are added. Graphs with the external quark self-energies are not shown in the Figure, but should be included in the 
calculation.}
\label{VirtualCorr}
\end{center}
\begin{center}
\includegraphics[scale=0.7]{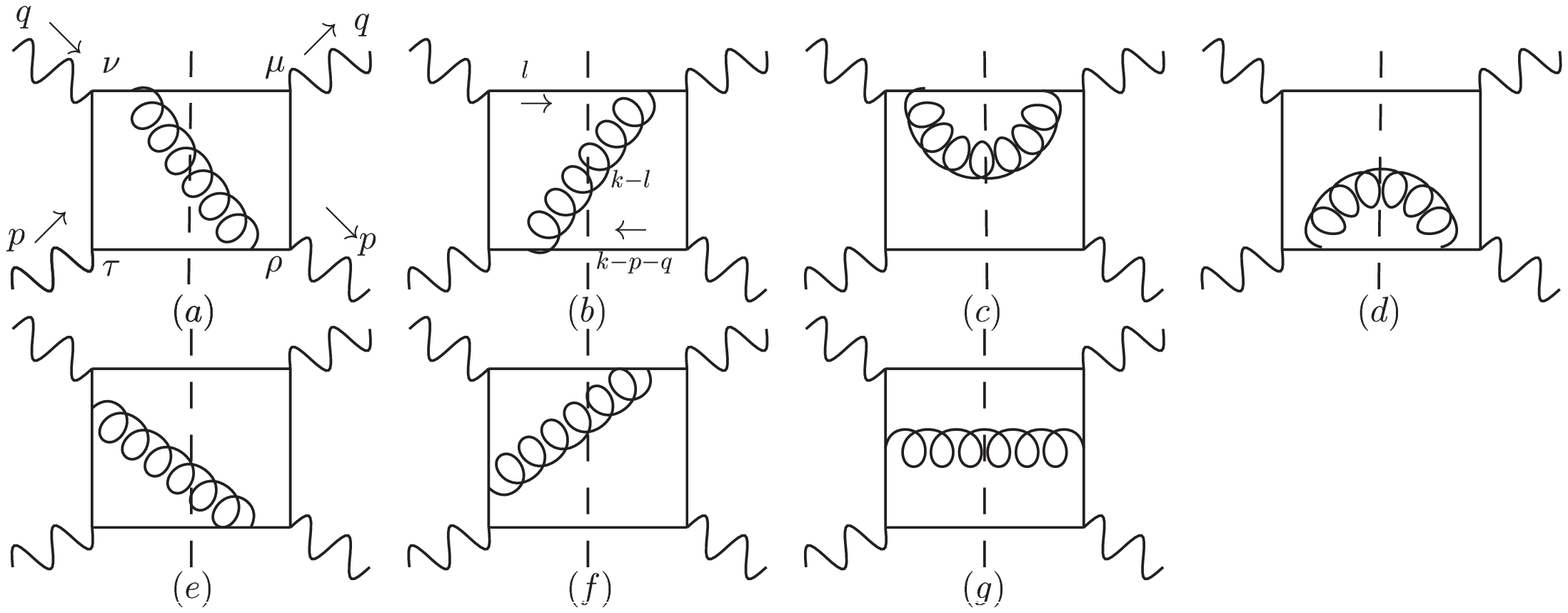}
\caption{Two-loop diagrams with a real gluon emission. Similar graphs corresponding to (e) and (f) are included. Also graphs with $(q, \mu)$ and $(p,\rho)$ interchanged are added. }
\label{RealEmis}
\end{center}
\end{figure}

We calculate the cross sections for 
the two photon annihilation to the heavy quark $q_H{\overline q}_H$ pairs 
\be
\gamma^*(q)+\gamma(p)\longrightarrow q_H+{\overline q}_H~, \label{BoxwithGluon}
\ee
with one-loop gluon corrections and to the gluon bremsstrahlung processes
\be
\gamma^*(q)+\gamma(p)\longrightarrow q_H+{\overline q}_H+g~.\label{Brems}
\ee
We employ the technique developed by Anastasiou and Melnikov \cite{AnaMel}, which is 
based on the Cutkosky rules  and the reduction of Feynman integrals to master integrals.
First, following the Cutkosky rules \cite{Cutkosky}, the delta-functions which appear 
in the phase space integrals are replaced with differences of two propagators
\be
2\pi i\delta(r^2-m^2)\rightarrow \frac{1}{r^2-m^2+i0}-\frac{1}{r^2-m^2-i0}~,\label{CutkoskyRule}
\ee
where $m$ is the heavy quark mass.
Then the cross sections for the virtual corrections to the processes (\ref{BoxwithGluon}) and for the  bremsstrahlung processes (\ref{Brems}) are described by the two-loop diagrams shown in 
Fig.\ref{VirtualCorr} and Fig.\ref{RealEmis}, respectively, where a cut propagator should be understood as the r.h.s. of Eq.(\ref{CutkoskyRule}).

We regularize the amplitudes by  dimensional regularization $D=4-2\epsilon$. Then we  apply the following $D$-dimensional projection operator
\be
 (P_{TT}^a)^{\mu\nu\rho\tau} = \frac{1}{2(D-2)(D-3)} (R^{\mu\rho}R^{\nu\tau}-R^{\mu\tau}R^{\nu\rho}) ~,
\ee
with
\be
R^{\mu\rho}=-g^{\mu\rho}+\frac{q^{\mu} p^{\rho}+q^{\rho} p^{\mu}}{p\cdot q}-\frac{q^2p^{\mu} p^{\rho} }{(p\cdot q)^2}~,
\ee
to these diagrams to extract the contributions to 
$g_1^\gamma$.  They are expressed in terms of a large number of two-loop scalar integrals of the form
\bea
&&\hspace{-1.5cm}A(\nu_i)
\equiv A(\nu_{k},\nu_{kq},\nu_{kp},\nu_{kpq},\nu_{l},\nu_{lq},\nu_{lp},\nu_{lpq},\nu_{kl})\nn\\
&&\hspace{-1cm}=\int\frac{d^D k}{(2\pi)^D}\frac{d^Dl}{(2\pi)^D}\frac{1}{\left[k^2-m^2 \right]^{\nu_{k}}\left[(k-q)^2 -m^2\right]^{\nu_{kq}}\left[(k-p)^2 -m^2\right]^{\nu_{kp}}\left[(k-p-q)^2 -m^2\right]^{\nu_{kpq}}}\nn\\
&&\times \frac{1}{\left[l^2-m^2 \right]^{\nu_{l}}\left[(l-q)^2 -m^2\right]^{\nu_{lq}}
\left[(l-p)^2 -m^2\right]^{\nu_{lp}}\left[(l-p-q)^2 -m^2\right]^{\nu_{lpq}}\left[(k-l)^2\right]^{\nu_{kl}}}~,\nn\\\label{As}
\eea
and the coefficients of these integrals are written as functions of $x, Q^2, m^2$ and $D$. 
Note that $1/(k-l)^2$ is a gluon propagator. 
Actually  $A(\nu_i)$ has seven propagators at most and thus at least two $\nu_i$'s are zero.

We arrange the integration variables $k$ and $l$ so that the cut propagators are
$1/[k^2-m^2]$ and $1/[(k-p-q)^2-m^2]$ for the diagrams in Fig.\ref{VirtualCorr}.
Among many $A(\nu_i)$s there appear those with one or both of the cut propagators 
eliminated. Those integrals do not contribute to 
$g_1^\gamma$. Thus we only pick up $A(\nu_i)$s which are in the form 
$A(1,\nu_{kq},\nu_{kp},1,\nu_{l},\nu_{lq},\nu_{lp},\nu_{lpq},\nu_{kl})$ and discard others.  A similar procedure is applied to the diagrams in  Fig.\ref{RealEmis}.
We choose $1/[l^2-m^2]$, $1/[(k-p-q)^2-m^2]$ and $1/(k-l)^2$ for the cut propagators and, therefore, search $A(\nu_i)$s in the form 
$A(\nu_{k},\nu_{kq},\nu_{kp},1,1,\nu_{lq},\nu_{lp},\nu_{lpq},1)$ and discard others.

The number of the relevant $A(\nu_i)$s is still large. Then, following the reduction procedure
\cite{Laporta} which is based on the method of integration by parts 
\cite{Tkachov} and the use of the Lorentz invariance of  scalar integrals~\cite{GehrmannRemiddi}, these 
$A(\nu_i)$s can be expressed in terms of fewer number of master integrals. 
Today several public codes~\cite{Anastasiou:2004vj,FIRE,Studerus:2009ye}  are available. We make use of FIRE  and  express the relevant $A(\nu_i)$s as a linear combination of the master integrals which are denoted as 
\be
M(\nu_i)\equiv M(\nu_{k},\nu_{kq},\nu_{kp},\nu_{kpq},\nu_{l},\nu_{lq},\nu_{lp},\nu_{lpq},\nu_{kl})~,
\ee
in the same way as the notation of $A(\nu_i)$s in Eq.(\ref{As}). Again the master integrals 
in the form of $M(1,\nu_{kq},\nu_{kp},1,\nu_{l},\nu_{lq},\nu_{lp},\nu_{lpq},\nu_{kl})$ are 
only relevant for the virtual correction diagrams in Fig.\ref{VirtualCorr} and 
those in the form of $M(\nu_{k},\nu_{kq},\nu_{kp},1,1,\nu_{lq},\nu_{lp},\nu_{lpq},1)$ are relevant 
for the real gluon emission diagrams in Fig.\ref{RealEmis}. 

Finally we perform the  phase space integrations  for these cut master integrals. 
For the two-cut and three-cut master integrals, we evaluate
\bea
&&\hspace{-1cm}{\rm Disc}^{(2)}~M(1,\nu_{kq},\nu_{kp},1,\nu_{l},\nu_{lq},\nu_{lp},\nu_{lpq},\nu_{kl})\nn\\
&&\hspace{-1cm}\equiv \int\frac{d^D k}{(2\pi)^{D}}(2\pi)\delta^{(+)}(k^2-m^2)(2\pi)\delta^{(+)}\left((p+q-k)^{2}-m^2\right)
\frac{1}{\left[(k-q)^2 -m^2\right]^{\nu_{kq}}\left[(k-p)^2 -m^2\right]^{\nu_{kp}}}\nn\\
&&\hspace{-0.5cm}\times\int\frac{d^Dl}{(2\pi)^D}\frac{1}{\left[l^2-m^2 \right]^{\nu_{l}}\left[(l-q)^2 -m^2\right]^{\nu_{lq}}
\left[(l-p)^2 -m^2\right]^{\nu_{lp}}\left[(l-p-q)^2 -m^2\right]^{\nu_{lpq}}\left[(k-l)^2\right]^{\nu_{kl}}}~,\nn\\
&&\label{PhaseSpaceInt2}\\
&&\hspace{-1cm}{\rm and}\nn\\
&&\hspace{-1cm}{\rm Disc}^{(3)}~M(\nu_{k},\nu_{kq},\nu_{kp},1,1,\nu_{lq},\nu_{lp},\nu_{lpq},1)\nn\\
&&\hspace{-1cm}\equiv \int \int\frac{d^D k}{(2\pi)^{D}}\int\frac{d^D l}{(2\pi)^{D}}~ 
(2\pi)\delta^+((k-l)^2)(2\pi)\delta^+(l^2-m^2)(2\pi)\delta^+((p+q-k)^2-m^2)\nn\\
&&\times\frac{1}{\left[k^2-m^2 \right]^{\nu_{k}}\left[(k-q)^2 -m^2\right]^{\nu_{kq}}\left[(k-p)^2 -m^2\right]^{\nu_{kp}}}\nn\\
&&\times \frac{1}{\left[(l-q)^2 -m^2\right]^{\nu_{lq}}
\left[(l-p)^2 -m^2\right]^{\nu_{lp}}\left[(l-p-q)^2 -m^2\right]^{\nu_{lpq}}}~,\label{PhaseSpaceInt3}
\eea
respectively. Note that  at least two 
$\nu_i$'s are zero in both (\ref{PhaseSpaceInt2}) and (\ref{PhaseSpaceInt3}).

There appear 61 master integrals in total in this  analysis of $g_1^\gamma(x,Q^2)$. However, the choice 
of a set of master integrals is not unique. We are at liberty to replace a master integral with one of the other scalar integrals. We choose a set of master integrals such that each corresponding coefficient function
is finite in the limit $D\rightarrow 4$~\cite{ChetyrkinFST}. With this choice of the set, 
the phase space integrations for  master integrals need only be evaluated  up to the finite terms in the series expansion in $\epsilon$. 

When the virtual correction diagrams in Fig.\ref{VirtualCorr} are concerned, 
the ultraviolet (UV) singularities appear in the graphs (b), (c) and (d), while 
the infrared (IR) singularity emerges from the graph (a). Both the UV and IR singularities are regularized 
by dimensional regularization. The UV singularities are removed by renormalization. We adopt the on-shell scheme both for the wave function renormalization of the external 
quark and the mass renormalization. For the latter, we  replace the bare mass in the Born cross section by the renormalized mass $m$, 
\be
m_{bare}\rightarrow m\left[1+\frac{\alpha_s}{4\pi}C_F S^{\epsilon}\left(
\frac{\mu^2}{m^2}\right)^{\epsilon}\left\{-\frac{3}{\epsilon}-4\right\}\right]~,
\ee
where $C_F=\frac{4}{3}$ is the Casimir factor, $S^{\epsilon}=(4\pi)^\epsilon e^{-\epsilon \gamma_E}$ with 
Euler constant $\gamma_E$ and $\mu$ is the arbitrary reference scale of dimensional regularization.The renormalization of the QCD gauge coupling constant is not 
necessary at this order.
 The IR singularities   appear also in the real gluon emission graphs (a),(b), (c) and (d) of Fig.\ref{RealEmis}. 
However, the IR singularities cancel when the both contributions from the virtual correction graphs and the real gluon
emission graphs are added. Actually the IR singularities reside in the two-cut master integrals in the form 
 $M(1,\nu_{kq},\nu_{kp},1,1,\nu_{lq},\nu_{lp},1,1)$ and the three-cut master integrals 
$M(\nu_{k},\nu_{kq},\nu_{kp},1,1,\nu_{lq},\nu_{lp},\nu_{lpq},1)$ with $\nu_{k}+\nu_{lpq}=2$.
The details of the calculation will be reported elsewhere \cite{WKS}.

\begin{figure}
\begin{center}
\includegraphics[scale=0.9]{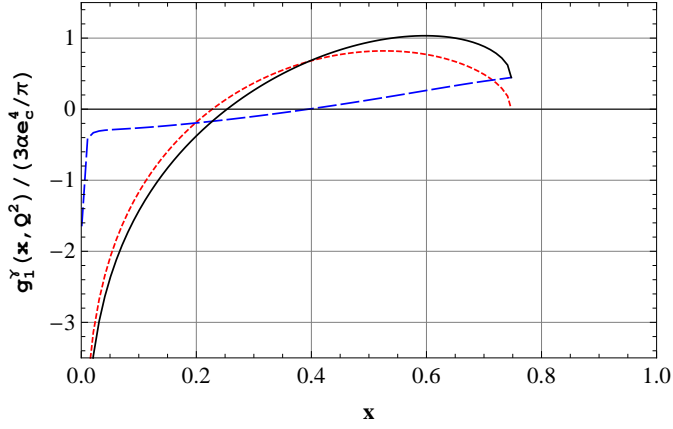}           
\caption{Charm quark effects on the polarized real photon structure function $g_1^\gamma (x,Q^2)$ in the PM in units of $(3\alpha e_c^4/\pi)$ for $Q^2=20~ {\rm GeV}^2$ and 
$m_c=1.3$ GeV with $\alpha_s=0.22$. We plot the LO result (dotted line), 
the NLO contribution (dashed line) and the sum of LO and NLO contributions (solid line).}
\label{g1Result}
\end{center}
\end{figure}

\begin{figure}
\begin{center}
\includegraphics[scale=0.9]{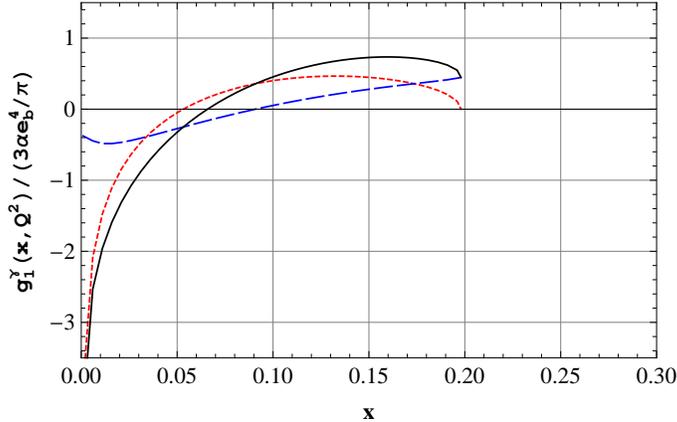}           
\caption{Bottom quark effects on the polarized real photon structure function $g_1^\gamma (x,Q^2)$ in the PM in units of $(3\alpha e_b^4/\pi)$ for $Q^2=20~ {\rm GeV}^2$ and 
$m_b=4.5$ GeV with $\alpha_s=0.22$. We plot the LO result (dotted line),
the NLO contribution (dashed line) and the sum of LO and NLO contributions (solid line).}
\label{g1ResultBottom}
\end{center}
\end{figure}

In Figs.\ref{g1Result} and \ref{g1ResultBottom} we plot the polarized real photon structure function $g_1^\gamma (x,Q^2)$ predicted by the massive PM up to the NLO for the case of $Q^2=20~ {\rm GeV}^2$ and  $\alpha_s=0.22$. We choose $c$ and $b$ as a 
heavy quark, for Figs. \ref{g1Result} and \ref{g1ResultBottom}, respectively.
We take $m_c=1.3~ {\rm GeV}, m_b=4.5 ~{\rm GeV}$, $e_c=\frac{2}{3}$ and $e_b=-\frac{1}{3}$. 
Here we show three curves: the LO result,  the sum of LO and NLO corrections and the NLO corrections alone.   The allowed $x$ region is $0\le x \le x_{\rm max}$ ~with  
\be
x_{\rm max}=\frac{1}{1+\frac{4m^2}{Q^2}}~.
\ee
The LO result is expressed by
\be
g_1^\gamma (x,Q^2)|_{\rm LO}=\frac{3\alpha}{\pi}e_q^4\left\{
\left( \ln\frac{1+\beta}{1-\beta}  \right)(2x-1)+\beta(-4x+3)
\right\}~,\label{g1PMLO}
\ee
where
\be
\beta=\sqrt{1-\frac{4m^2 x}{Q^2(1-x)}}~.
\ee
For $x\rightarrow x_{\rm max}$, $\beta$ goes to zero and thus $g_1^\gamma (x,Q^2)_{\rm LO}$ vanishes at $x_{\rm max}$.

We observe in the Figures that there exist NLO corrections both at large and small $x$, positive at large $x$ but negative in small $x$ region, a behavior similar to the LO result. Especially, the radiative corrections are large near the threshold 
(near $x_{\rm max}$) and the NLO curve does not vanish at $x_{\rm max}$. This is due to the well-known Coulomb 
singularity, which appears when the Coulomb gluon is exchanged between the quark and anti-quark pair near threshold. The diagram Fig.\ref{VirtualCorr}(a) is responsible for this threshold behavior. 
The virtual correction  to the left of the cut line in Fig.\ref{VirtualCorr}(a) gives rise to a factor  $1/\beta$ while 
a factor $\beta$ comes out from the phase space integration. They are combined and yield a finite  but non-zero result at $x_{\rm max}$. 

We consider the sum rule (\ref{SumRuleg1gammaReal}) for a real photon target. Substituting the LO result 
$g_1^\gamma (x,Q^2)|_{\rm LO}$ given by (\ref{g1PMLO}), we see the sum rule holds 
\cite{ET, BASS,NSV,FS,BBS}.  Figs.\ref{g1Result} and \ref{g1ResultBottom} show  that the sum rule also seems to be satisfied by the NLO contribution to $g_1^\gamma$ in both cases. Expressing the NLO contribution as 
$g_1^\gamma (x,Q^2)|_{\rm NLO}$, we find numerically
\be
\int_0^{x_{\rm max}}g_1^\gamma (x,Q^2)|_{\rm NLO} dx=0~.
\ee 
But due to the limitation of accuracy of our numerical integration, we observed
\bea
\delta\equiv\frac{\int_0^{x_{\rm max}}g_1^\gamma (x,Q^2)|_{\rm NLO} dx}
{\int_0^{x_{\rm max}}\Big|g_1^\gamma (x,Q^2)|_{\rm NLO}\Big| dx}=
\left(-2.2,~ -2.0\right) \times 10^{-4}~.\nn
\eea
for charm and bottom cases, respectively.
So we conclude that the sum rule is satisfied in the massive PM up to the NLO. 

However, if we go on further and analyse $g_1^\gamma (x,Q^2)$ to higher orders in 
perturbation theory, we expect that the result will diverge at $x_{\rm max}$ due to the 
Coulomb singularity.  A detail analysis on the structure of the Coulomb singularity
tells that $g_1^\gamma |_{\rm NNLO}\sim \beta \times \left( \alpha_s/\beta\right)^2$~\cite{HSY,Kiyo} whose
integral for the first moment is ill-defined due to end-point singularity at $x=x_{\rm max}$.  
The sum rule is not well-defined in the perturbation theory starting at NNLO.
To obtain an appropriate threshold behavior for photon structure functions,
we may resort to the method of resummation of the Coulomb singularities.
A noticeable difference in the resummation is  emergence of bound-state poles 
of $q_H{\overline q}_H$  above $x_{\rm max}$.
Then the left-hand side of the 
sum rule Eq. (\ref{SumRuleg1gammaReal})  should include also the 
bound-state contributions. We will not pursue this issue further here 
but render it to our future publications.

In summary we have calculated the NLO corrections to the polarized 
photon structure function $g_1^\gamma$ in the massive PM. We have found that 
the NLO contribution is noticeable at large $x$ and does not vanish at 
$x_{\rm max}$ due to the Coulomb singularity. 
We have also found numerically that the sum rule  (\ref{SumRuleg1gammaReal}) 
is satisfied up to the NLO in the massive PM. The details of our calculation 
will be reported elsewhere~\cite{WKS}.

\vspace{2cm}

\vspace{0.5cm}
\leftline{\large\bf Acknowledgement}
\vspace{0.5cm}

We thank T. Uematsu and T. Ueda for valuable discussions. At the early stage T. Ueda 
participated in this investigation.

\clearpage





\begin{thebibliography}{99}

\bibitem{LHC} http://lhc.web.cern.ch/lhc/.

\bibitem{ILC} http://www.linearcollider.org/cms/. 

\bibitem{KS}
         K.~Sasaki, {\sl Phys. Rev.} {\bf D22} (1980) 2143; 
          {\sl Prog. Theor. Phys. Suppl. }~{\bf 77} (1983) 197.


\bibitem{SV}
         M. Stratmann and W. Vogelsang, {\it Phys. Lett.} {\bf B386}, 
         (1996) 370 .

\bibitem{SmithvanNeerven}
          E.~Laenen, S.~Riemersma, J.~Smith and W.L.~van Neerven,  {\sl Phys.~Rev.} {\bf D49} (1994) 5753; W.~Beenakker, H.~Kuijf, W.L.~van Neerven and J.~Smith,  {\sl Phys.~Rev.} {\bf D40} (1989) 54.

\bibitem{GRS}
          M.~Gl{\"u}ck, E.~Reya and C.~Sieg, {\sl Phys.~Lett.}
         {\bf B503}, 285 (2001);\\  {\sl Eur. Phys. J.} {\bf C20}, 271 (2001).


\bibitem{Cutkosky}
      R.E.~Cutkosky, {\sl J.~Math.~Phys.} {\bf 1} (1960) 429.

\bibitem{ET}
        A.~V.~Efremov and O.~V.~Teryaev, JINR Report NO. E2-88-287, Dubna, 
        1988; {\sl Phys.~Lett.} {\bf B240}, 200 (1990).
\bibitem{BASS}
          S.~D.~Bass, {\sl Int.~J.~Mod.~Phys.} {\bf A7}, 6039 (1992).
\bibitem{NSV}
          S.~Narison, G.~M.~Shore and G.~Veneziano,
           {\sl Nucl.~Phys.} {\bf B391}, 69 (1993); \\
          G.~M.~Shore and G.~Veneziano, {\sl Mod.~Phys.~Lett.}
           {\bf A8}, 373 (1993);\\
          G.~M.~Shore and G.~Veneziano, {\sl Nucl.~Phys.} {\bf B381}, 23 (1992);\\
          G.~M.~Shore; {\sl Nucl.~Phys.} {\bf B712}, 411 (2005).


\bibitem{FS}
          A.~Freund and L.~M.~Sehgal, {\sl Phys.~Lett.} {\bf B341}, 90 (1994).
\bibitem{BBS}
          S.~D.~Bass, S.~J.~Brodsky and I.~Schmidt,
           {\sl Phys.~Lett.} {\bf B437}, 424 (1998).

\bibitem{DHG}
       S. D. Drell and A. C. Hearn, {\sl Phys. Rev. Lett.}{\bf 162} (1966) 1520; 
      S. B. Gerasimov, {\sl Yad.~Fiz.} {\bf 2} (1965) 839.


\bibitem{AnaMel}
     C.~Anastasiou and K.~Melnikov, {\sl Nucl.~Phys.} {\bf B646} (2002) 220.

\bibitem{Laporta}
     S.~Laporta, {\sl Int.~J.~Mod.~Phys.} {\bf A15} (2000) 5087.


\bibitem{Tkachov}
      F.V.~Tkachov, {\sl Phys.~Lett.} {\bf B100} (1981) 65;
      K.G.~Chetyrkin and F.V.~Tkachov, {\sl Nucl.~Phys.} {\bf B192} (1981) 159.

\bibitem{GehrmannRemiddi}
     T.~Gehrmann and E.~Remiddi, {\sl Nucl.~Phys.} {\bf B580} (2000) 485.

\bibitem{Anastasiou:2004vj}
  C.~Anastasiou, A.~Lazopoulos,
  {\sl JHEP}  {\bf 0407 } (2004)  046.

\bibitem{FIRE}
     A.V. Smirnov, {\sl JHEP} 0810:(2008)107

\bibitem{Studerus:2009ye}
  C.~Studerus,
  {\sl Comput.\ Phys.\ Commun.\ }  {\bf 181 } (2010)  1293-1300.


\bibitem{ChetyrkinFST}
     K.G.~Chetyrkin, M.~Faisst, C.~Sturm and M.~Tentyukov, {\sl Nucl.~Phys.} {\bf B742} (2006) 208.

\bibitem{WKS}
     N.~Watanabe, Y.~Kiyo and K.~Sasaki, in preparation.

\bibitem{HSY}
     K.~Hagiwara, Y.~Sumino and H.~Yokoya, {\sl Phys.~Lett.} {\bf B666} (2008) 71. 

\bibitem{Kiyo}
     Y.~Kiyo, J.H.~K\"{u}hn, S.~Moch, M.~Steinhauser and P. Uwer, {\sl Eur. Phys. J.} {\bf C60} (2009) 375.



\end{thebibliography}
\end{document}